\documentclass{style/sig-alternate}
\usepackage{lmodern}
\usepackage[utf8]{inputenc}
\usepackage[TS1,T1]{fontenc}
\usepackage[usenames,dvipsnames,svgnames,table]{xcolor}
\usepackage{url}
\usepackage{graphicx}
\DeclareGraphicsExtensions{.eps}
\usepackage{latexsym}
\usepackage{xspace}
\usepackage{enumitem}
\usepackage{subfigure}
\usepackage{balance}
\usepackage{algorithm}
\usepackage{algorithmic}
\usepackage{graphicx}

\setenumerate{noitemsep,topsep=0pt,parsep=0pt,partopsep=0pt}

\newcommand*{\eat}[1]{}

\newcommand*{\kl}{\textsf{rKL}\xspace}
\newcommand*{\nd}{\textsf{rND}\xspace}
\newcommand*{\rd}{\textsf{rRD}\xspace}

\def\btau{\boldsymbol{\tau}}
\def\bsigma{\boldsymbol{\sigma}}

\begin{document}

\title{Measuring Fairness in Ranked Outputs}

\author{
   Ke Yang \\
  Drexel University \\
  ky323@drexel.edu  \and 
  Julia Stoyanovich\thanks{This work was supported in part by NSF Grants No. 1464327 and 1539856, and BSF Grant No. 2014391.} \\
  Drexel University \\
  stoyanovich@drexel.edu
}

\maketitle

\thispagestyle{empty}

\begin{abstract}

Ranking and scoring are ubiquitous.  We consider the setting in which
an institution, called a ranker, evaluates a set of individuals based
on demographic, behavioral or other characteristics.  The final output
is a ranking that represents the relative quality of the individuals.
While automatic and therefore seemingly objective, rankers can, and
often do, discriminate against individuals and systematically
disadvantage members of protected groups.  This warrants a careful
study of the fairness of a ranking scheme.

In this paper we propose fairness measures for ranked outputs.  We
develop a data generation procedure that allows us to systematically
control the degree of unfairness in the output, and study the behavior
of our measures on these datasets.  We then apply our proposed
measures to several real datasets, and demonstrate cases of
unfairness.  Finally, we show preliminary results of incorporating our
ranked fairness measures into an optimization framework, and show
potential for improving fairness of ranked outputs while maintaining
accuracy.

\end{abstract}

\section{Introduction}
\label{sec:intro}

Ranking and scoring are ubiquitous.  We consider the setting in which
an institution, called a ranker, evaluates a set of items (typically
individuals, but also colleges, restaurants, websites, products, etc)
based on demographic, behavioral or other characteristics.  The output
is a ranking representing the relative quality of the items.

Rankings are used as the basis of important decisions including college 
admissions, hiring, promotion, grant making, and lending.  As a result, 
rankings have a potentially enormous impact on the livelihood and well-being 
of individuals. While automatic and therefore seemingly objective, 
rankers can discriminate against individuals and systematically
disadvantage members of protected groups~\cite{CitronP14}. This warrants a 
careful study of the {\em fairness of a ranking scheme}, a topic we investigate 
in this paper.

A useful dichotomy is between individual fairness --- a requirement that 
similar individuals are treated similarly, and  group fairness, also known 
as statistical parity --- a requirement that demographics of those receiving 
a particular {\em positive outcome} are identical to the demographics of the 
population as a whole~\cite{DBLP:conf/innovations/DworkHPRZ12}. The focus of this paper
is on statistical parity, and in particular on the case where a subset of the population 
belongs to a {\em protected  group}.  Members of a protected group
share a characteristic that cannot be targeted for discrimination.  In the US, 
such characteristics include race, gender and disability status, among others.

We make a simplifying assumption and consider membership in one protected group at a time (i.e., 
 we consider only gender or disability status or membership in a minority group).  Further,
we assume that membership in a protected group is binary (minority vs. majority group, 
rather than a break-down by ethnicity). 
This is a reasonable starting point, and is in line with much recent literature~\cite{DBLP:conf/innovations/DworkHPRZ12,DBLP:conf/icml/ZemelWSPD13,DBLP:journals/corr/Zliobaite15a}.

In this paper we propose several measures that quantify statistical
parity, or lack thereof, in ranked outputs.  The reasons to consider
this problem are two-fold.  First, having insight into the properties of
a ranked output, or more generally of an output of 
an algorithmic process, helps make the process transparent, interpretable and accountable~\cite{Kroll2017,StoyanovichG2016}. Second, principled fairness quantification
mechanisms allow us to engineer better algorithmic rankers, correcting for bias
in the input data that is due to the effects of historical discrimination against 
members of protected groups.

To reason about fair assignment of outcomes to groups, we must first understand the 
meaning of a {\em positive outcome} in our context. 
There is a large body of work on measuring discrimination in machine learning,
see~\cite{DBLP:journals/corr/Zliobaite15a} for a survey. 
The most commonly considered algorithmic task is binary classification, where items assigned to 
the positive class are assumed to receive the positive outcome. In contrast to classification, 
a ranking is, by definition, relative, and so outcomes for items being ranked are not strictly 
positive or negative.  The outcome for an item ranked among the top-5 is at least as good 
as the outcome for an item ranked among the top-10, which is in turn at least as good as 
the outcome for an item ranked among the top-20, etc.

{\bf Basic idea.} Our formulation of fairness measures is based on the following 
intuition: Because it  is more beneficial for an item to be ranked higher, it 
is also more important to achieve statistical parity at higher ranks. The idea 
is to take several well-known statistical parity measures proposed in the literature~\cite{DBLP:journals/corr/Zliobaite15a}, and make them rank-aware 
by placing them within the nDCG framework~\cite{DBLP:journals/tois/JarvelinK02}.
nDCG is commonly used for evaluating the quality of ranked lists 
in information retrieval, and is appropriate here because it provides a 
principled weighted discount mechanism.  Specifically, we calculate 
{\em set-wise parity} at a series of cut-off points in a 
ranking, and progressively sum these values with a position-based discount.  

\eat{We are inspired by several of the recently proposed discrimination criteria,
but observe that a ranked outcomes pose new challenges.
This is due to two
reasons.  The first reason is that a ranking is, by definition,
relative.  Consider score-based rankers, which compute a score
independently for each item and then sort items in order of decreasing
score. Even for these relatively simple rankers (which are in contrast
to, e.g., learning to rank
methods~\cite{DBLP:series/synthesis/2014Li}) the final outcome --- the
position of an item in the ranked output --- cannot be determined
based solely on the item's score, but rather is due to the
relationship between an item's score and the scores of other items in
the dataset.  That is, an outcome is computed w.r.t. to a set of
items, not for each item individually.  The second reason is that
ranked outcomes are not strictly positive or negative, as is the case
with, e.g., binary classification.}

The rest of this paper is organized as follows.  We fix our notation
and describe a procedure for generating synthetic rankings of varying
degrees of fairness in Section~\ref{sec:prelim}.  In
Section~\ref{sec:measures} we present novel ranked fairness measures
and show their behavior on synthetic and real datasets. In
Section~\ref{sec:opt} we show that fairness in rankings can be
improved through an optimization framework, but that additional work
is needed to integrate the fairness and accuracy measures into the
framework more tightly, to achieve better performance.  We conclude in
Section~\ref{sec:conc}.

\section{Preliminaries}
\label{sec:prelim}

We are given a database $I (\underline{k}, s, x_1, \ldots, x_m)$ of
items.  In this database, an item identified by $k$ is associated with
an attribute $s$ that denotes membership in a binary protected group,
and with descriptive attributes $x_1, \ldots, x_m$.  We denote by
$S^{+} \subseteq I$ the items that belong to the protected group, and
by $S^{-} = I \setminus S^{+}$ the remaining items.\eat{  We use $N$ to
denote the size of $I$.}

\begin{figure}
\centering
\includegraphics[width=1.8in]{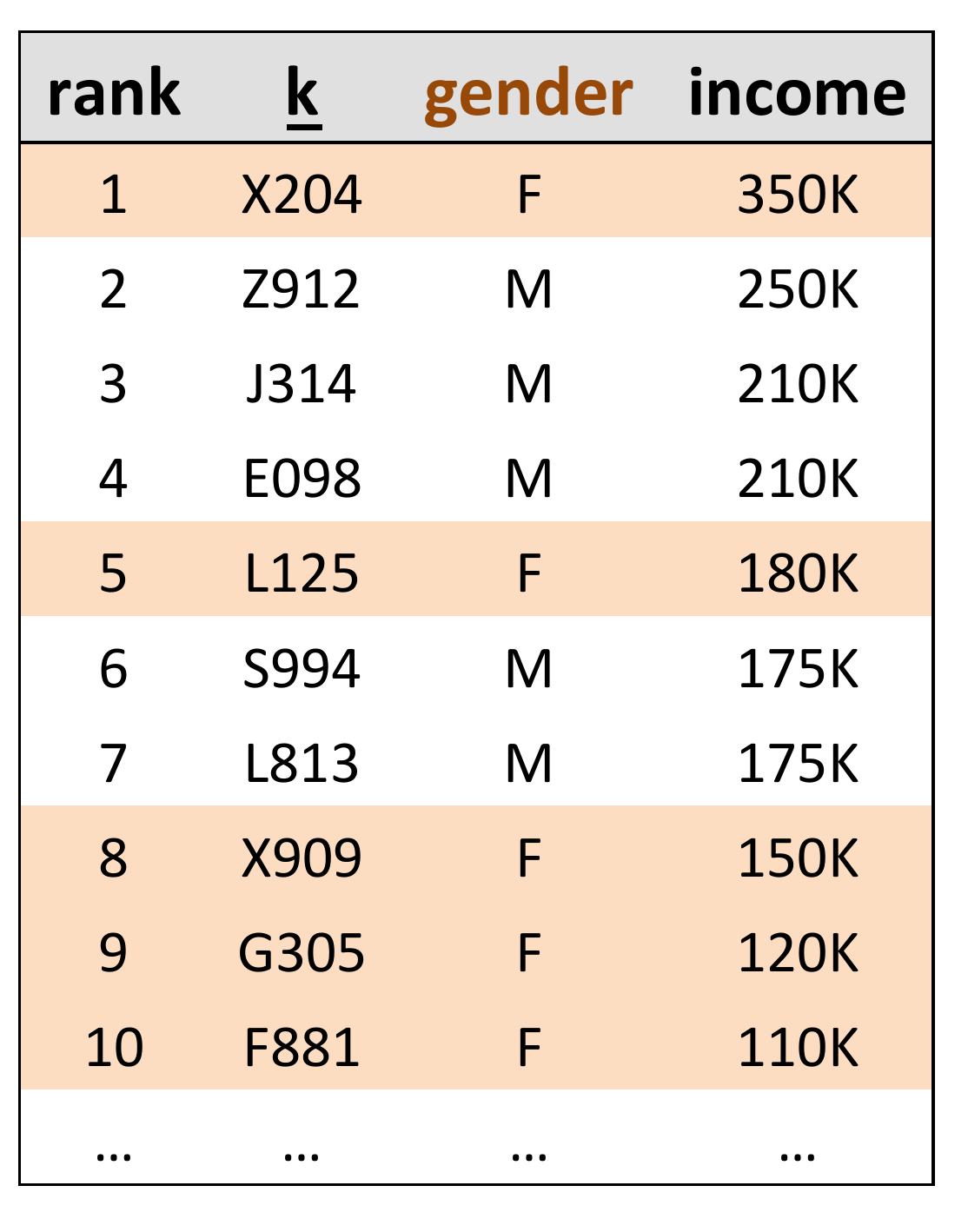}
\caption{A ranked list, sorted in descending order of income, $gender=F$ is protected.}
\label{fig:ex}
\end{figure}

A {\em ranking} over $I$ is a bijection $\btau: I \rightarrow \{1,
\ldots, N \}$, where $\btau(k)$ is the position of the item with
identifier $k$ in $\btau$.  We refer to the item at position $i$ in
$\btau$ by $\tau_{i}$.  For a pair of items $k_1, k_2$, we say that
$k_1$ is {\em preferred} to $k_2$ in $\btau$, denoted $k_1
\succ_{\btau} k_2$ if $\btau(k_1) < \btau(k_2)$. Figure~\ref{fig:ex}
gives an example of rankings of individuals, with gender as the
protected attribute.

We denote by $\btau_{1..i}$ the top-$i$ portion of $\btau$, and by
$S^{+}_{1 \ldots i}$ the items from the protected group $S^{+}$ that
appear among the top-$i$: $S^{+} \cap \btau_{1..i}$.

\begin{figure}
\centering
\includegraphics[width=3in]{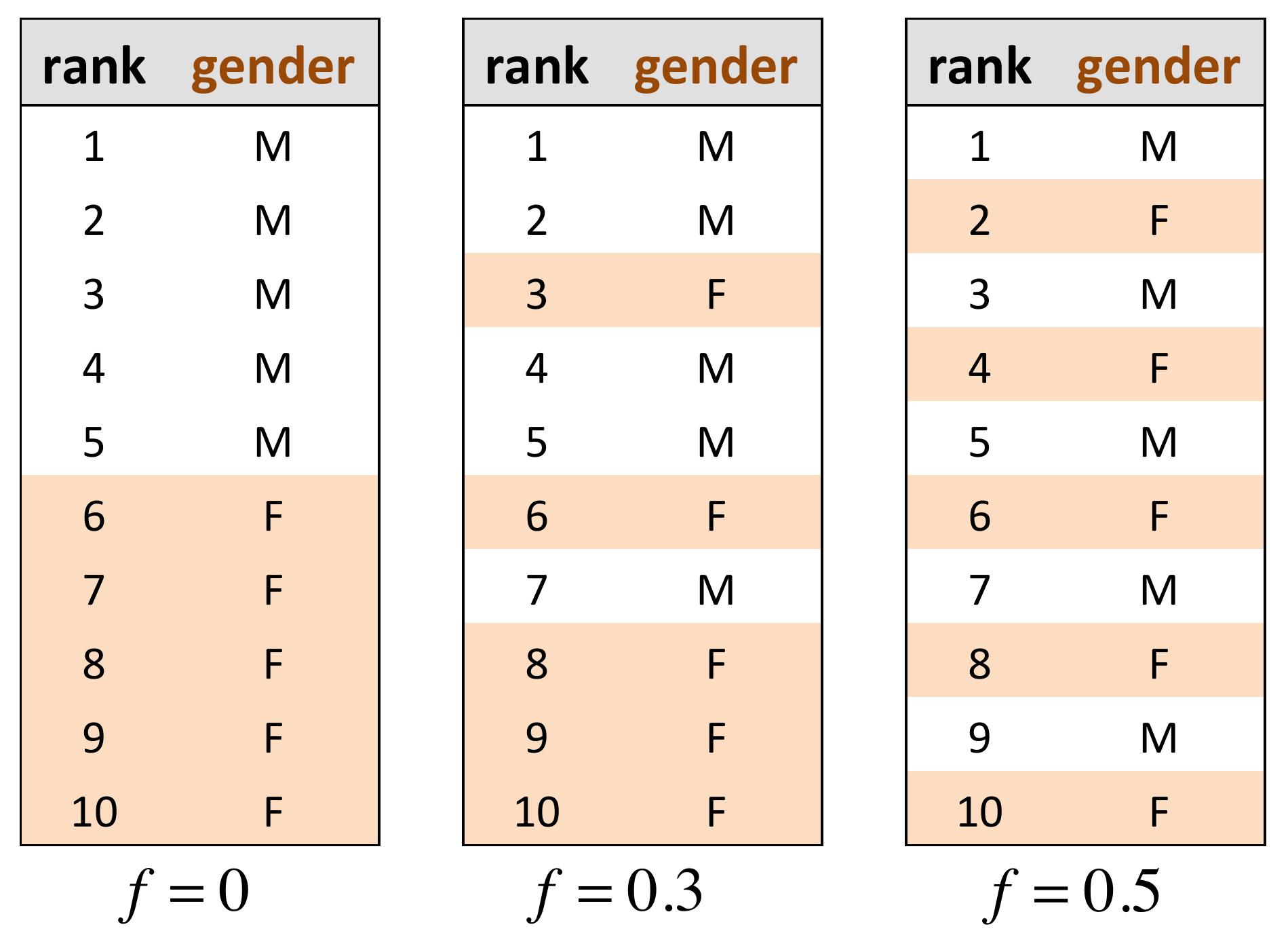}
\caption{Three ranked lists, $gender=F$ is the protected group.}
\label{fig:ex3}
\end{figure}

\subsection*{Data generator}

To systematically study the behavior of the proposed measures, we
generate synthetic ranked datasets of varying degrees of fairness.
Algorithm~\ref{alg:rankgen} presents our data generation procedure.
This algorithm takes two inputs.  The first is a ranking $\btau$,
which may be a random permutation of the items in $I$, or it may be
generated by the vendor according to their usual process, e.g., in a
score-based manner.  The second input is the fairness probability $f \in
[0, 1]$, which specifies the relative preference between items in
$S^{+}$ and in $S^{-}$.  When $f = 0.5$, groups $S^{+}$ and $S^{-}$
are mixed in equal proportion for as long as there are items in both
groups.  When $f > 0.5$, members of $S^{+}$ are preferred, and when $f
< 0.5$ members of $S^{-}$ are preferred.  In extreme cases, when $f =
0$, all members of $S^{-}$ will be placed in the output ranking $\bsigma$ before any
members of $S^{+}$ (first all male individuals, then all female in
Figure~\ref{fig:ex}); when $f = 1$, the opposite will be true: all of
$S^{+}$ will be ranked higher than $S^{-}$ in the output $\bsigma$.  

The following additional property holds for a pair $k_1, k_2 \in I$: if
$(k_1.s = k_2.s) \wedge (k_1 \succ_{\btau} k_2)$ then $k_1
\succ_{\bsigma} k_2$. That is, Algorithm~\ref{alg:rankgen} does not
change the relative order among a pair of items that are both in
$S^{+}$ or in $S^{-}$.

\begin{algorithm}[h!]
\caption{Unfair ranking generator}
\begin{algorithmic}[1]
\REQUIRE Ranking $\btau$, fairness probability $f$.\\
\COMMENT {Initialize the output ranking $\bsigma$.}\\
\STATE $\bsigma \leftarrow \emptyset$\\
\STATE $\btau^{+} = \btau \cap S^{+}$\\
\STATE $\btau^{-} = \btau \cap S^{-}$\\
\WHILE {$(\btau^{+} \neq \emptyset) \wedge (\btau^{-} \neq \emptyset)$}
\STATE $p=random([0, 1])$\\
\IF{$p<f$}
\STATE $\bsigma \leftarrow pop(\btau^{+})$\\
\ELSE
\STATE $\bsigma \leftarrow pop(\btau^{-})$\\
\ENDIF
\ENDWHILE\\
\STATE $\bsigma \leftarrow \btau^{+}$\\
\STATE $\bsigma \leftarrow \btau^{-}$\\
\RETURN $\bsigma$\\
\end{algorithmic}
\label{alg:rankgen}
\end{algorithm}

Figure~\ref{fig:ex3} gives examples of three ranked lists of 10 individuals, 
with the specified fairness probabilities, and with $gender=F$ as the protected group.  
For $f=0$, all male individuals are placed ahead of the females.  For $f=0.3$, males and 
females are mixed, but males occur more frequently at top ranks.  For $f=0.5$, males and 
females are mixed in equal proportion.

\section{Fairness measures}
\label{sec:measures}

\begin{figure*}[t!]
\centering
\begin{minipage}{2.2in}
\centering
\includegraphics[width=2.2in]{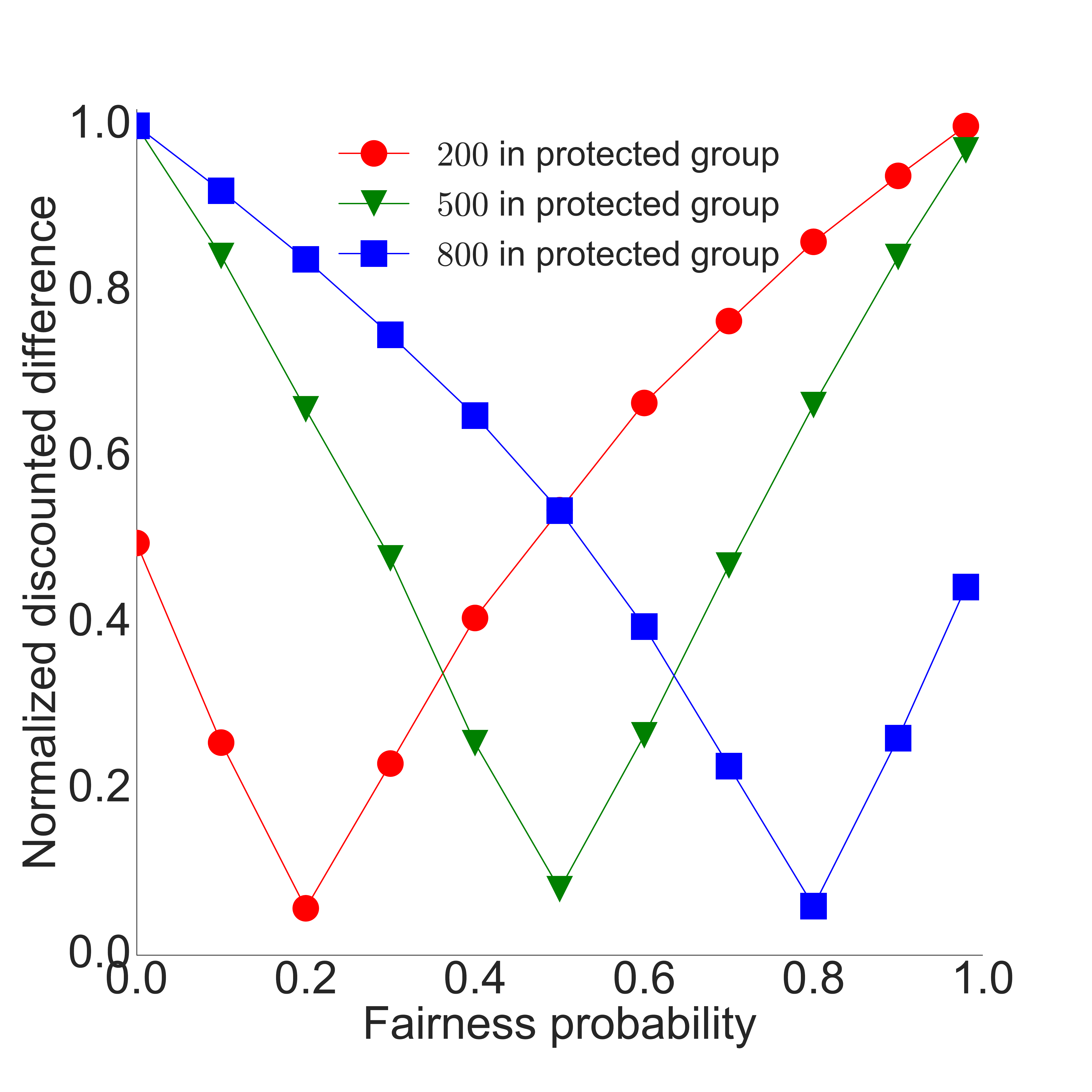}
\caption{\nd on 1,000 items}
\label{fig:nd}
\end{minipage}
\begin{minipage}{2.2in}
\centering
\includegraphics[width=2.2in]{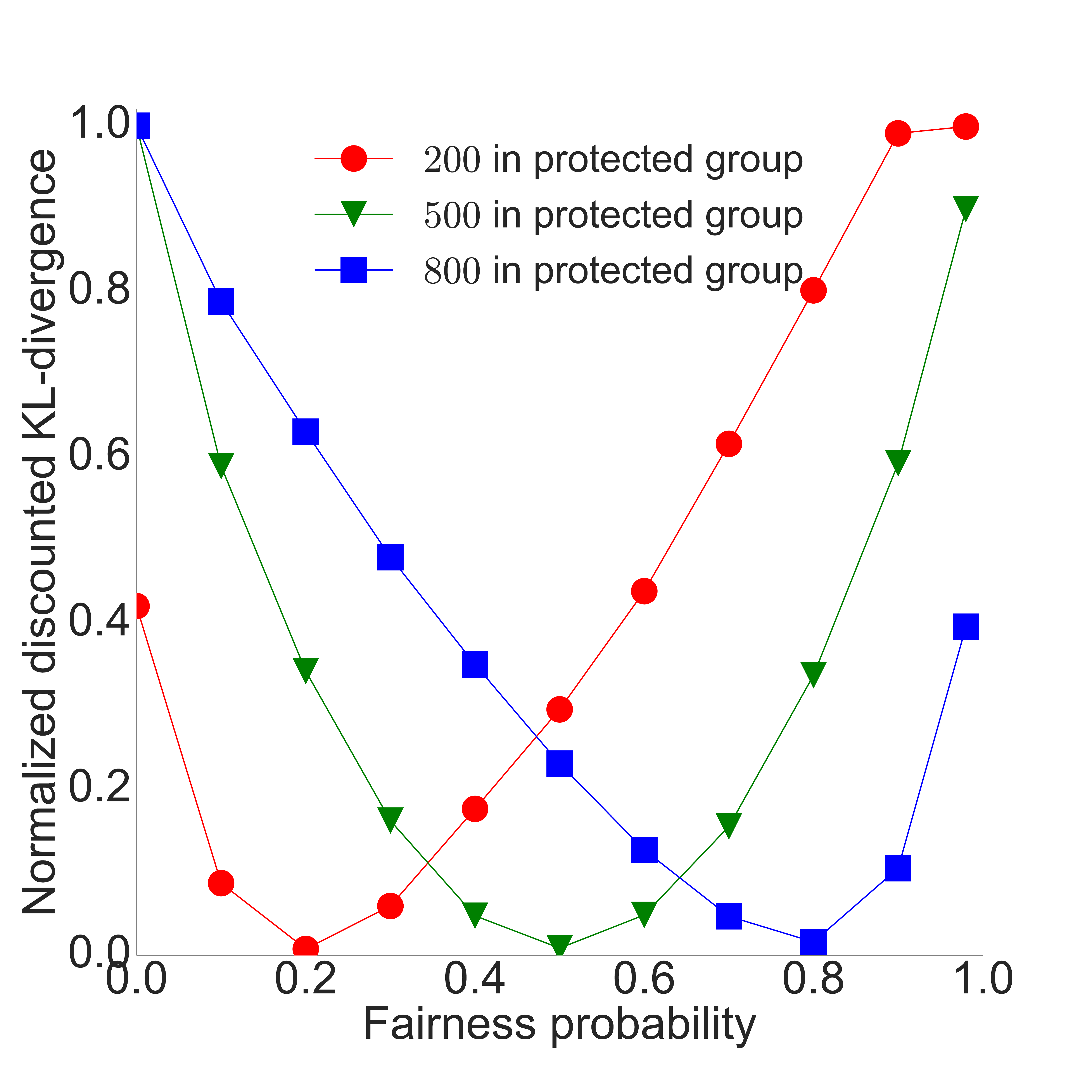}
\caption{\kl on 1,000 items}
\label{fig:kl}
\end{minipage}
\begin{minipage}{2.2in}
\centering
\includegraphics[width=2.2in]{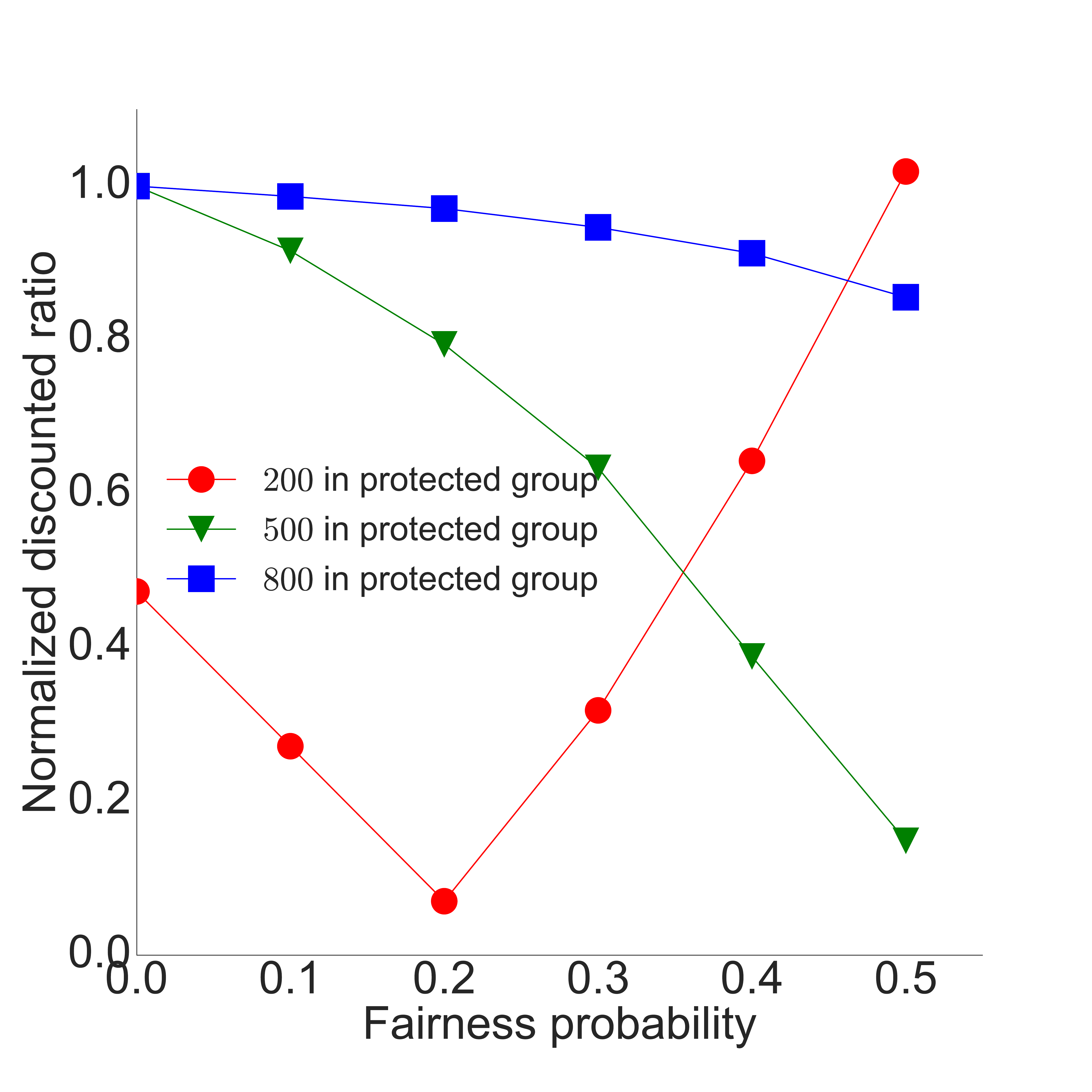}
\caption{\rd on 1,000 items}
\label{fig:rd}
\end{minipage}
\end{figure*}

A ranking scheme exhibits statistical parity if membership in a
protected group does not influence an item's position in the output.
We now present three measures of statistical parity that capture this
intuition.

Our measures quantify the relative representation of the protected
group $S^{+}$ in a ranking $\btau$.  For all measures, we compute
{\em set-based fairness} at discrete points in the ranking (top-$10$, top-$20$, etc),
and compound these values with a logarithmic discount.  In this way, we
express that higher positions in the ranking are more important, i.e.,
that it is more important to be fair at the top-$10$ than at the
top-$100$.  Our logarithmic discounting method is inspired by that used
in nDCG~\cite{DBLP:journals/tois/JarvelinK02}, a ranked quality
measure in Information Retrieval.

All measures presented in this section are normalized to the $[0,1]$
range for ease of interpretation.  All measures have their best (most
fair) value at 0, and their worst value at 1.

\subsection*{Normalized discounted difference (\nd)}

Normalized discounted difference (\nd) (Equation~\ref{eq:nd}),
computes the difference in the proportion of members of the protected
group at top-$i$ and in the over-all population.  

Values are
accumulated at discrete points in the ranking with a logarithmic
discount, and finally normalized.  Normalizer $Z$ is computed as the
highest possible value of \nd for the given number of items $N$ and
protected group size $|S^{+}|$.

\begin{equation}
\nd (\btau)=\frac{1}{Z} \sum_{i=10,20,...}^{N}{ \frac{1}{log_{2}{i}} \left|\frac{|S^{+}_{1 \ldots i}|}{i} - \frac{|S^{+}|}{N} \right|}
\label{eq:nd}
\end{equation}

Figure~\ref{fig:nd} plots the behavior of \nd on synthetic datasets of
1000 items, with 200, 500 and 800 items in $S^{+}$, as a function of
fairness probability.  We make four observations: (1) Groups $S^{+}$
and $S^{-}$ are treated symmetrically --- a low proportion of either
$S^{+}$ or $S^{-}$ at high ranks leads to a high (unfair) \nd score.
(2) The best (lowest) value of \nd is achieved when fairness parameter
is set to the value matching the proportion of $S^{+}$ in $I$: 0.2 for
200 protected group members out of 1000, 0.5 for 500 members, and 0.8
for 800 members. (3) \nd is convex and continuous.  (4) \nd is not
differentiable at 0.

We argued in the introduction that fairness measures are important
not only as a means to observe properties of the output, but also
because they allow us to engineer processes that are more fair.
A common approach is to specify an optimization problem in which 
some notion of accuracy or utility is traded against some notion of 
fairness~\cite{DBLP:conf/icml/ZemelWSPD13}.  While \nd presented above is convex and 
continuous, it is not differentiable, limiting its usefulness in an optimization
framework.  This consideration motivates us to develop an alternative measure,
\kl, described next.

\subsection*{Normalized discounted KL-divergence (\kl)}

Kullback-Leibler (KL) divergence measures the expectation of the
logarithmic difference between two discrete probability distributions
$P$ and $Q$:

\begin{equation}
D_{KL}(P||Q) = \sum_{i}P(i) log \frac{P(i)}{Q(i)}
\label{eq:klpq}
\end{equation}

We use KL-divergence to compute the expectation of the difference
between protected group membership at top-$i$ vs. in the over-all
population.  We take:

\begin{equation}
 P = \left(\frac{|S^{+}_{1 \ldots i}|}{i},\frac{|S^{-}_{1 \ldots i}|}{i} \right),  Q = \left(\frac{|S^{+}|}{N}, \frac{|S^{-}|}{N} \right)
\label{eq:p}
\end{equation}

and define normalized discounted KL-divergence (\kl) as:

\begin{equation}
\kl (\btau)=\frac{1}{Z} \sum_{i=10,20,...}^{N}{ \frac{D_{KL}(P||Q)}{log_{2}{i}}}
\label{eq:kl}
\end{equation}

\eat{\begin{equation}
\kl=\frac{1}{Z}\sum_{i=10,20,...}^{N}{\frac{\frac{S^{1}_{i}}{i}log(\frac{S^{1}_{i}}{i}\frac{N}{S_{N}^{1}})+\frac{\ S^{0}_{i}}{i}log(\frac{S^{0}_{i}}{i}\frac{N}{S_{N}^{0}})}{log_{2}{i}}}
\label{eq:kl}
\end{equation}}

Figure~\ref{fig:kl} plots the behavior of \kl on synthetic datasets of
1000 items, with 200, 500 and 800 items in $S^{+}$, as a function of
fairness probability.  We observe that this measure has similar
behavior as \nd (Equation~\ref{eq:nd} and Figure~\ref{eq:nd}), but
that it appears smoother, and so may be more convenient to optimize
robustly.  An additional advantage of \kl is that it can be used
without modification to go beyond binary protected group membership,
e.g., to capture proportions of different racial groups, or age
groups, in a population.

\eat{\begin{figure}[t!]
\centering
\includegraphics[width=2.8in]{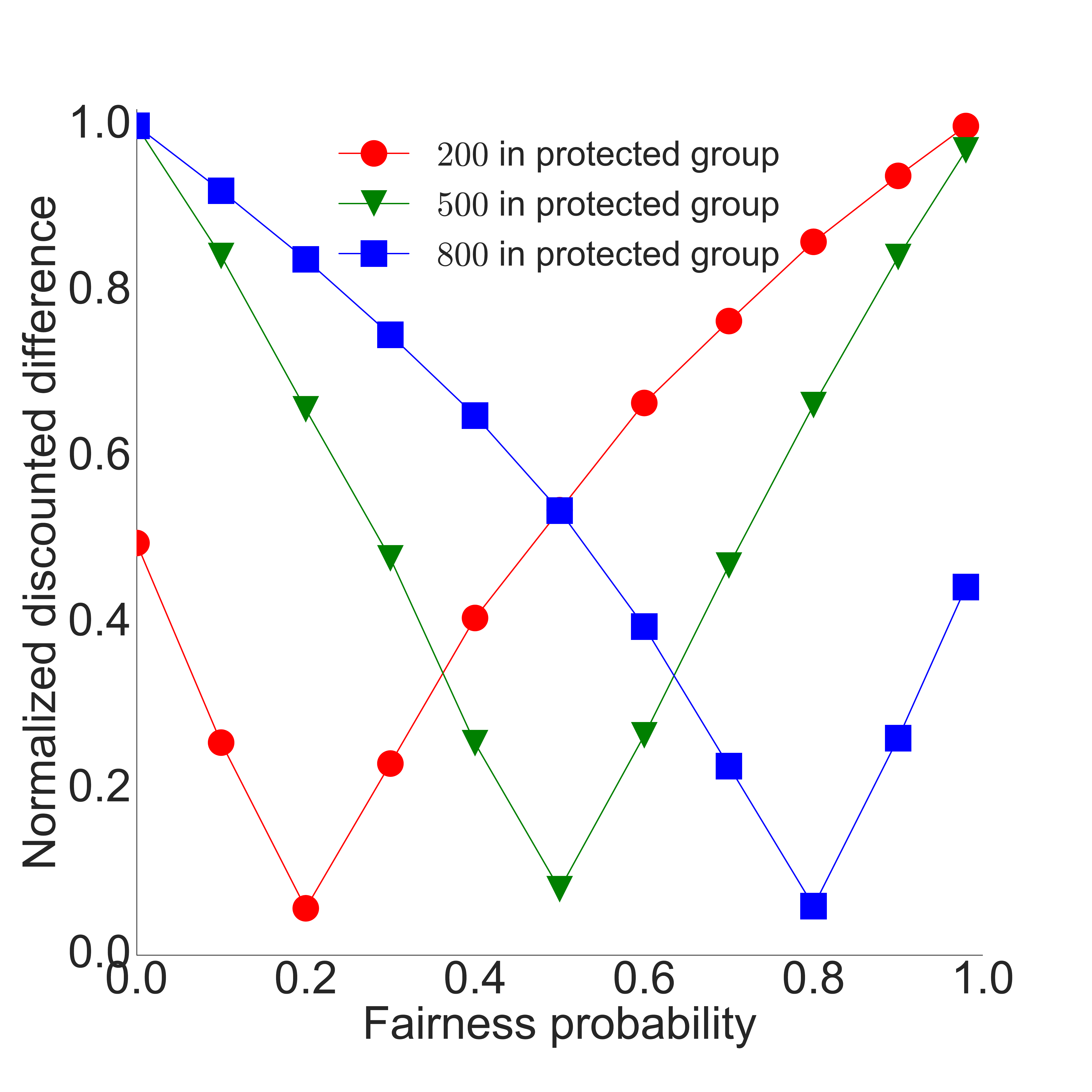}
\caption{Behavior of normalized discounted difference (\nd) on a
  synthetic dataset with 1000 items.}
\label{fig:nd}
\end{figure}}

\eat{\begin{figure}
\centering
\includegraphics[width=2.8in]{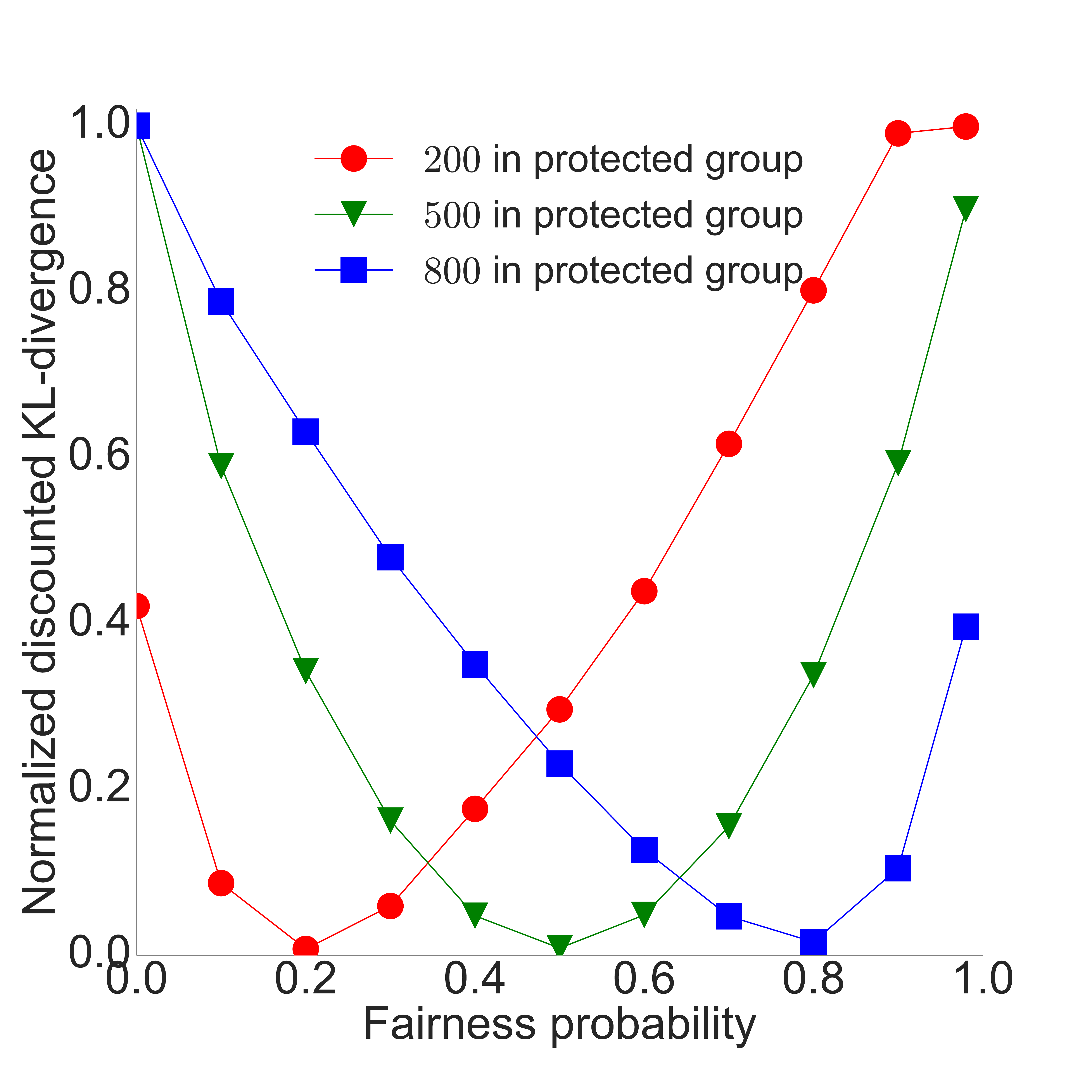}
\caption{Behavior of normalized discounted KL-divergence (\kl) on synthetic data.}
\label{fig:kl}
\end{figure}}

\eat{\begin{figure}[b!]
\centering
\includegraphics[width=3in]{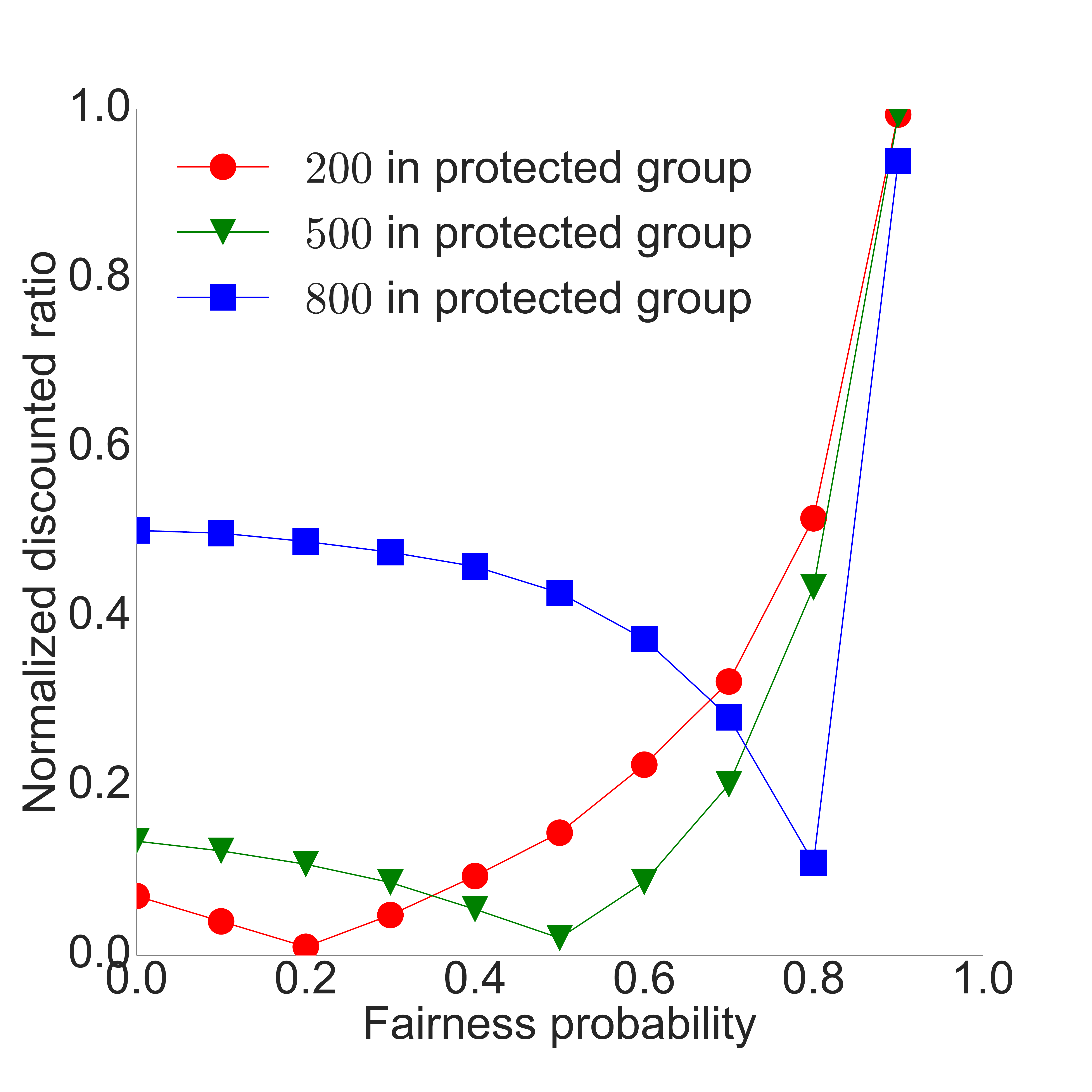}
\caption{Behavior of normalized discounted ratio (\rd) on synthetic data.}
\label{fig:rd}
\end{figure}}

\subsection*{Normalized discounted ratio (\rd)}

Our final measure, normalized discounted ratio (\rd), is formulated
similarly to \nd, with the difference in the denominator of the
fractions: the size of $S^{+}_{1 \ldots i}$ is compared to the size of
$S^{-}_{1 \ldots i}$, not to $i$ (and similarly for the second term,
$S^{+}$).  When either the numerator or the denominator of a fraction
is 0, we set the value of the fraction to 0.

Behavior of the \rd measure on a synthetic dataset is presented in
Figure~\ref{fig:rd}.  We observe that \rd reaches its best (lowest)
value at the same points as do \nd and \kl, but that it shows
different trends.  Most importantly, because \rd does not treat
$S^{+}$ and $S^{-}$ symmetrically, its behavior when protected group
represents the majority of the over-all population (800 protected group members out of 1000
in Figure~\ref{fig:rd}), or when $S^{+}$ is preferred to $S^{-}$ (fairness probability $>0.5$), 
is not meaningful.
We conclude that {\em \rd is
  only applicable when the protected group is the minority group},
i.e., when $S^{+}$ corresponds to at most 50\% of the underlying
population, and when fairness probability is below 0.5. 

\begin{equation}
\rd (\btau)=\frac{1}{Z} \sum_{i=10,20,...}^{N}{ \frac{1}{log_{2}{i}} \left|\frac{|S^{+}_{1 \ldots i}|}{|S^{-}_{1 \ldots i}|} - \frac{|S^{+}|}{|S^{-}|} \right|}
\label{eq:rd}
\end{equation}

\eat{\begin{equation}
\rd=\frac{1}{Z}\sum_{i=10,20,...}^{N}{\frac{|S^{+}_{1 \ldots i}/S^{0}_{i}-S^{1}_{N}/S^{0}_{N}|}{log_{2}{i}}}
\label{eq:rd}
\end{equation}}

\begin{figure*}[t!]
\centering
\includegraphics[width=6.5in]{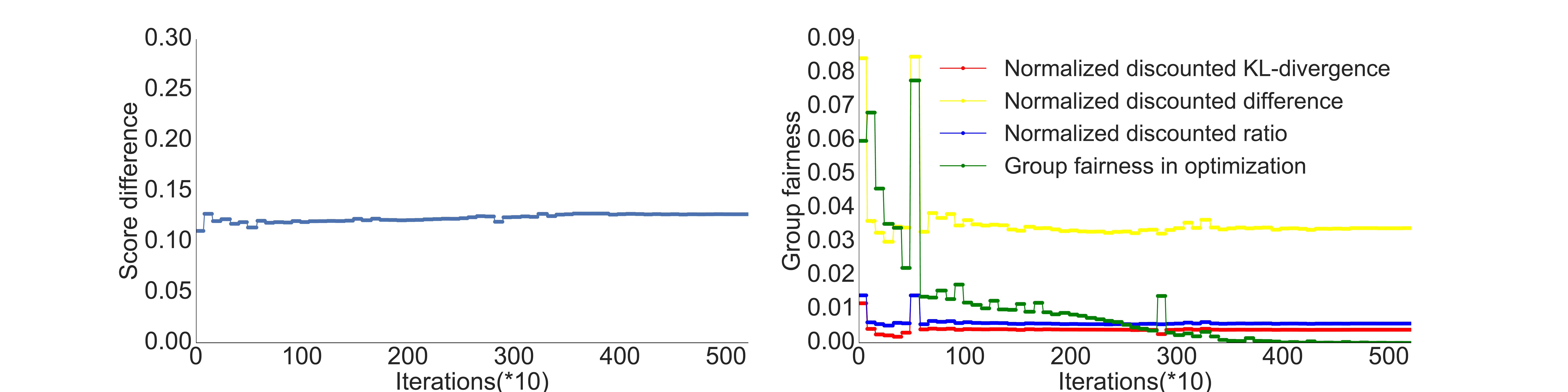}
\caption{Accuracy and fairness on German Credit, ranked by {\em sum of
    normalized attribute values}, with $k=10$.}
\label{fig:weighted}
\end{figure*}

\begin{figure*}
\centering
\includegraphics[width=6.5in]{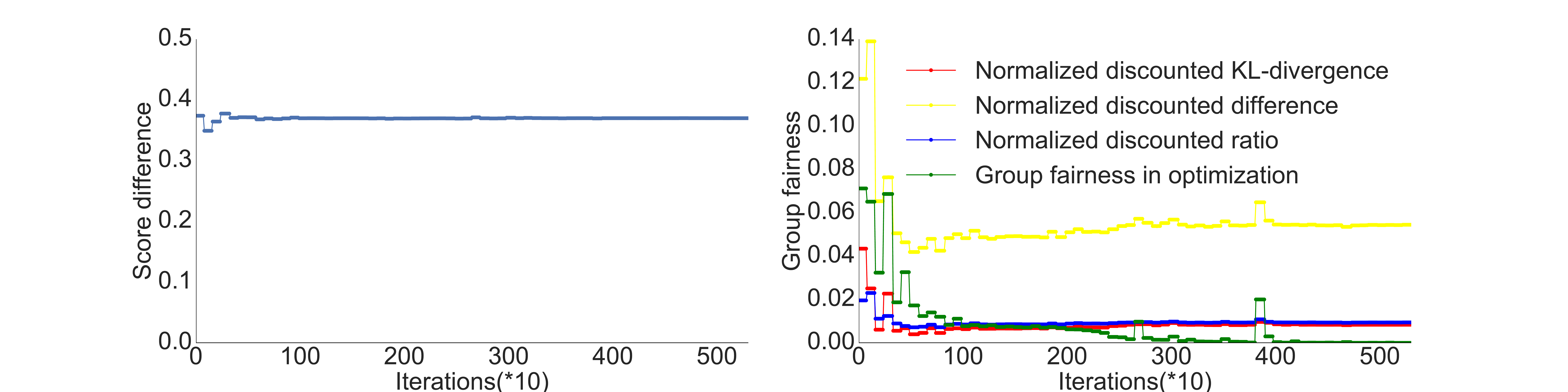}
\caption{Accuracy and fairness on German Credit, ranked by {\em credit
    amount}, with $k=10$.}
\label{fig:credit}
\end{figure*}

\subsection*{Evaluation with real datasets}

We used several real datasets to study the relative behavior of our
measures, and to see whether our measures signal any fairness
violations in that data.  We highlight results on two datasets:
ProPublica~\cite{propublica} and German Credit~\cite{Lichman:2013}.
Results on other datasets, including CS Rankings~\cite{CSrankings}, US
News \& World Report rankings
(\url{premium.usnews.com/best-colleges}), and Adult
Income~\cite{Lichman:2013} show interesting trends but we do not
include them due to space limitations.  These will be discussed in the
full version of the paper.

{\bf ProPublica} is the recidivism data analyzed by~\cite{propublica}
and retrieved
  from~\url{https://github.com/propublica/compas-analysis}.  This
dataset contains close to 7,000 criminal defendant records.  The goal
of the analysis done by~\cite{propublica} was to establish whether
there is racial bias in the software that computes predictions of
future criminal activity. Racial bias was indeed ascertained, as was
gender bias, see~\cite{propublica} for details.  In our analysis we
set out to check whether ranking on the values of recidivism score,
violent recidivism score, and on the number of prior arrests shows
parity w.r.t. race (black as the protected group, 51\% of the dataset)
and gender (female as the protected group, 19\% of the dataset).

Using race as the protected attribute, we found \nd = 0.44
for ranking on recidivism and \nd
= 0.44. We found lower but still noticeable \kl values for these
rankings (0.17 and 0.18, respectively).  Interestingly, ranking by the
number of prior arrests produced \nd = 0.23 but \kl = 0.04, showing
much higher unfairness according to the \nd measure than to the \kl.
Note that \rd is inapplicable here, since the protected group
constitutes the majority of the population.

Using gender as the protected attribute, we found \nd = 0.15 for
ranking on recidivism score, \nd = 0.12 for violent recidivism and \nd
= 0.11 for ranking on prior arrests.  The \kl was low --- between 0.01 
and 0.02 for these cases.  We measured \rd = 0.20 for recidivism ranking,
 \rd = 0.14 for violent recidivism and \rd = 0.16 for prior arrests.

{\bf German Credit} is a dataset from~\cite{Lichman:2013} with
financial information about 1,000 individuals applying for loans.
This dataset is typically used for classification tasks.  Nonetheless,
several of the attributes that are part of this dataset can be used
for ranking, including duration (month), credit amount, status of
existing account, and employment length.  We experimented with ranking
on individual attributes duration (months) and credit amount, and also
used a score-based ranker that computes the score of each individual
by normalizing the value of each attribute, and then summing them with
equal weights.  We used all attributes that are either continuous or
discrete but ordered.

As protected attributes, we used gender (69\% female) and age (15\%
younger than 25, 55\% younger than 35).  For these cases, \kl ranged
between 0.01 and 0.15, and \nd ranged between 0.05 and 0.41.  \rd was
only applicable to age below 25, and ranged between 0.08 and 0.12.
We will show in Section~\ref{sec:opt} results of optimizing fairness
for this dataset, with $age < 25$ as the protected attribute.

\section{Learning fair rankings}
\label{sec:opt}

In this section we describe an optimization method for improving the
fairness of ranked outputs.  Our approach is based closely on the fair
representations framework of Zemel
et. al~\cite{DBLP:conf/icml/ZemelWSPD13}, which focuses on making
classification tasks fair.  We first briefly describe their framework,
and then explain our modifications that make it applicable to
rankings.

The main idea in~\cite{DBLP:conf/icml/ZemelWSPD13} is to introduce a
level of indirection between the input space $X$ that represents
individuals and the output space $Y$ in which classification outcomes
are assigned to individuals.  Specifically, they introduce a
multinomial random variable $Z$ of size $k$, with each of the $k$
values representing a ``prototype'' (cluster) in the space of $X$.
The goal is to learn a mapping from $X$ to $Z$ that preserves
information about attributes that are useful for the classification
task at hand, while at the same time hiding information about
protected group membership of $x$.  Statistical parity in this
framework is formulated as $P(Z=k~|~x \in S^{+}) = P(Z=k~|~x \in
S^{-}), \forall k$.

The goal is to learn a mapping that satisfies statistical parity and
at the same time preserves utility.  For this, $Z$ must be a good
description of $X$ (distances from $x\in X$ to its representation in
$\hat{x}$ should be small) and predictions based on $\hat{x}$ should
be accurate.  This formulation gives rise to the following
multi-criteria optimization problem, with the loss function $L = A_x
L_x + A_y L_y + A_z L_z$, where $L_x$, $L_y$ and $L_z$ are loss
functions and $A_x$, $A_y$, $A_z$ are hyper-parameters governing the
trade-offs.  We keep the terms responsible for statistical parity
($L_z$) and distance in the input space ($L_x$) as in the original
framework, and redefine the term that represents accuracy ($L_y$) as
appropriate for ranked outcomes.  We show here results that use
average per-item score difference between the ground-truth ranking
$\btau$ and its estimate $\hat{\btau}$ to quantify accuracy.  We also
experimented with position accuracy (per-item rank difference),
Kendall-$\tau$ distance, and Spearman and Pearson's correlation
coefficients, but omit these results due to space limitations, and
also because more work is needed to understand convergence properties
of the optimization under these measures.

Results of our preliminary evaluation on the German Credit dataset
(see Section~\ref{sec:measures}), with $age < 25$ as the protected
attribute, are presented in Figures~\ref{fig:weighted} (ranked on sum
of all attributes, normalized, with equal weights)
and~\ref{fig:credit} (ranked on the credit amount attribute).  We show
convergence of accuracy (normalized average score difference) and
fairness (\nd, \kl, \rd, and the statistical parity component $L_z$ as
``group fairness in optimization'').  We observe that all fairness
measures converge to a low (fair) value, but that accuracy is
optimized more effectively in Figure~\ref{fig:weighted} than in
Figure~\ref{fig:credit}, where the score difference is considerable.

\section{Conclusions and future work}
\label{sec:conc}

\balance
In this paper we presented novel measures that can be used to quantify
statistical parity in rankings.  We evaluated these measures in
synthetic and real datasets, and showed preliminary optimization
results.  Much future work remains on the evaluation of these measures
--- understanding their applicability in real settings, formally
establishing their mathematical properties, incorporating them into an
optimization framework, and ensuring that the framework is able to
both preserve accuracy and improve fairness.  One of the bottlenecks
in this process is the running time, which we are working to optimize
both by looking for ways to make fairness measures more
computationally friendly and by careful engineering.

\bibliographystyle{abbrv}
\bibliography{dataresp}

\end{document}